\documentclass[a4paper,twoside]{article}
\usepackage[strings]{underscore}

\usepackage{epsfig}
\usepackage{subcaption}
\usepackage{calc}
\usepackage{amssymb}
\usepackage{amstext}
\usepackage{amsmath}
\usepackage{amsthm}
\usepackage{multicol}
\usepackage{pslatex}
\usepackage{apalike}
\usepackage{url}
\usepackage{algorithm2e}
\usepackage[bottom]{footmisc}
\usepackage{SCITEPRESS}
\usepackage{graphicx}
\usepackage{xcolor}
\usepackage{booktabs}
\usepackage[most]{tcolorbox}
\usepackage[font=footnotesize,labelfont=bf]{caption}
\usepackage{todonotes}
\usepackage{hyperref}

\begin{document}

\title{KGiRAG: An Iterative GraphRAG Approach for Responding Sensemaking Queries}

\author{\authorname{Isabela Iacob, Melisa Marian and Gheorghe Cosmin Silaghi\orcidAuthor{0000-0002-3447-4736, corresponding author}}
\affiliation{Babe\c{s}-Bolyai University, Business Informatics Research Center, Cluj-Napoca, Romania}
\email{iacobisabela726@yahoo.com, melisa\_marian@yahoo.fr, gheorghe.silaghi@ubbcluj.ro}
}

\keywords{Knowledge Graphs, Large Language Models, RAG, Hallucination, Query Answering}

\abstract{Recent literature highlights the potential of graph-based approaches within large language model (LLM) retrieval-augmented generation (RAG) pipelines for answering queries of varying complexity, particularly those that fall outside the LLM’s prior knowledge. However, LLMs are prone to hallucination and often face technical limitations in handling contexts large enough to ground complex queries effectively. To address these challenges, we propose a novel iterative, feedback-driven GraphRAG architecture that leverages response quality assessment to iteratively refine outputs until a sound, well-grounded response is produced. Evaluating our approach with queries from the HotPotQA dataset, we demonstrate that this iterative RAG strategy yields responses with higher semantic quality and improved relevance compared to a single-shot baseline.
}

\onecolumn \maketitle \normalsize \setcounter{footnote}{0} \vfill

\section{\uppercase{Introduction}}
\label{sec:introduction}

As a recent advancement in Artificial Intelligence (AI), Large Language Models (LLMs) have demonstrated remarkable capabilities in solving a wide range of problems, including those within specialized domains such as medicine and business. They are increasingly being integrated into industrial production pipelines, and users worldwide rely on them to support various daily tasks. However, LLMs are also known to hallucinate in certain circumstances \cite{kandpal,hallucination}, particularly when required to generate responses grounded in large-scale, domain-specific, or proprietary knowledge \cite{hallucination}.

Retrieval-Augmented Generation (RAG) \cite{lewis2021} was developed to enable LLMs to answer queries based on data that is too large to fit within the model’s context window and was not encountered during training.
These queries can range from simple factual or multiple-choice questions to more complex ones that require paragraph-level responses and the synthesis of information drawn from extensive datasets.
The latter type — known as sensemaking queries — poses challenges for classical vector-based RAG systems \cite{edge2025}, as their limited context capacity constrains the integration of multiple, interrelated concepts necessary for coherent and comprehensive responses.
To address this limitation, GraphRAG approaches \cite{graphrag,zhangsurvey} have been introduced.
These methods leverage structured knowledge stored in graph-based repositories. During the retrieval phase, relevant facts are extracted from the graph, and the LLM generates responses grounded in this structured information.

While several approaches \cite{edge2025,grag,graphgpt} have demonstrated the feasibility of combining graph-structured knowledge with LLMs for addressing sensemaking queries, there remains significant room for improvement due to the intrinsic tendency of LLMs to hallucinate.
This paper contributes to this field by proposing an iterative, feedback-driven GraphRAG architecture based on Knowledge Graphs (KGs) designed to mitigate potential LLM hallucinations and to produce well-grounded, trustworthy responses
In contrast to static, one-shot methods such as that of Edge et al. \cite{edge2025}, our proposed approach introduces a feedback loop mechanism that evaluates the quality of each generated response. If the response does not meet a predefined quality threshold, the GraphRAG process is repeated with an expanded context retrieved from the KG.

To comprehensively evaluate the performance of the proposed architecture, we conduct experiments with queries from the public HotPotQA dataset \cite{hotpotqa}, testing multiple configurations of our feedback-based pipeline. The objective is to analyze how iterative retrieval and reasoning mechanisms affect answer quality, faithfulness, and contextual grounding, and to determine whether this looping architecture yields measurable improvements over a single-shot GraphRAG implementation.

The paper evolves as follows. In section \ref{seq:related} we revise related work about answering sensemaking queries and GraphRAG. 
Section \ref{seq:methodology} presents the architecture of our system for supporting the iterative feedback guided GraphRAG response generation. 
Section \ref{seq:experiments} describes our experimental approach. Section \ref{seq:results} presents and discusses the results. Section \ref{seq:conclusion} concludes the paper, reinforcing our main contributions. 

\section{\uppercase{Related Work}}
\label{seq:related}

Retrieval-Augmented Generation (RAG)~\cite{lewis2021} is a hybrid framework that integrates a retrieval component with a generative language model to enable grounded, knowledge-intensive question answering. Given an input query, the retriever searches through
a large-scale knowledge source to identify and return the most semantically relevant text fragments. These are subsequently provided as an evidence to the language model, which synthesizes a final answer grounded in these external sources.


Recent work, such as GraphGPT~\cite{graphgpt}, has demonstrated the value of integrating LLMs with graph-structured data through a graph instruction tuning paradigm. GraphGPT enables LLMs to understand complex graph structures and to perform effectively in both supervised and zero-shot settings, showcasing their reasoning capabilities when working with knowledge graphs. GRAG \cite{grag} shows that while graph-augmented retrieval delivers clear advantages for problems requiring multi-step reasoning and interconnected evidence, it may underperform for single-hop queries or when the graph quality or connectivity is poor.
Microsoft GraphRAG \cite{edge2025} addresses specifically the problem of responding sensemaking queries with a RAG approach. In the indexing phase they turn the corpus of documents into a KG, which is subsequently partitioned into hierarchical communities using a community detection algorithm. Within the retrieval phase, the relevant LLM-generated community summaries are  aggregated to produce the final response. \cite{graphrag} presents a survey of graph-based RAG approaches for constructing LLM-based query answering systems.

While LLMs are known to hallucinate \cite{hallucination}, sometimes they generate unsupported or misleading content when responding difficult questions. To improve the factual accuracy and coherence of responses, the RARR framework \cite{rarr} introduces an iterative refinement stage where the model revisits retrieved passages to fact-check and revise its initial generation. In the context of grounding the RAG process on a KG, the KiRAG approach \cite{kirag} performs an iterative retrieval process to dynamically identify and retrieve knowledge that bridges the information gaps of the LLM. 


Based on the extensive body of literature demonstrating the effectiveness of GraphRAG~\cite{graphrag,zhangsurvey}, our paper contributes by introducing a novel iterative, feedback-driven GraphRAG framework for addressing sensemaking queries. Assuming the existence of a domain-specific knowledge graph (KG), our method supplies the LLM with relevant context extracted from the KG. To automatically mitigate hallucination, we assess the quality of each generated response and, if it fails to meet predefined quality thresholds, the RAG process is repeated with an expanded context extracted from the KG. While inspired by the single-shot RAG approach of \cite{edge2025}, our method differs fundamentally in that it adopts an iterative refinement strategy.

\section{\uppercase{Methodology}}
\label{seq:methodology}

In this section we present our iterative feedback-based approach for GraphRAG. 

\begin{figure*}[!t]
  \centering
  \includegraphics[width=0.9\linewidth]{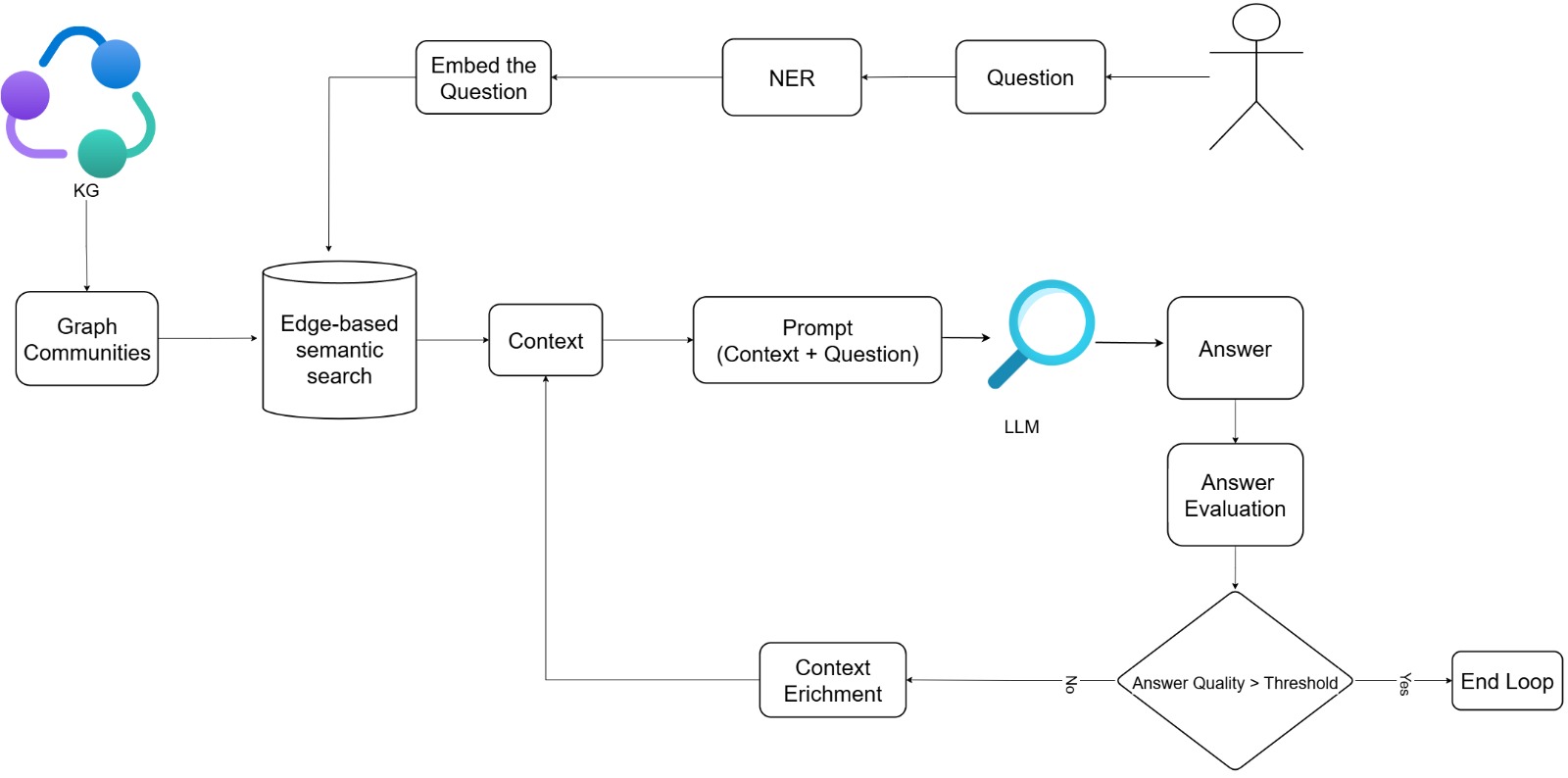}
  \caption{System architecture }
  \label{fig:arhitectura-ner}
\end{figure*}

\begin{figure}[!tb]
\scriptsize
\begin{tabular}{|p{7cm}|}
\hline
1. for each question: \\
2. \quad extract entities using NER \\
3. \quad embed the question for semantic search \\
4. \quad initialize context using entity-based semantic search \\
5. \quad repeat up to 4  times: \\
6. \quad \quad if first iteration: \\
7. \quad \quad \quad  retrieve initial evidence from graph \\
8. \quad \quad else: \\
9. \quad \quad \quad  expand graph with new related text units \\
10.\quad \quad update context via semantic search \\
11.\quad \quad generate answer with LLM using (question + context) \\
12.\quad \quad evaluate answer quality \\
13.\quad \quad if answer fulfills quality standards: \\
14.\quad \quad \quad            stop \\
\hline
\end{tabular}
\caption{KGiRAG processing pseudocode}%
\label{fig:pseudocode}%
\end{figure}

The proposed architecture presented in figures~\ref{fig:arhitectura-ner} and \ref{fig:pseudocode} extends and modifies the standard RAG framework by the following:
\begin{itemize}
\item we assume that a comprehensive KG encompassing all relevant domain-specific information is available. This KG serves as the main source of knowledge for the LLM-based retrieval process and is queried during the retrieval stage of the RAG framework. The KG is preprocessed with the community detection procedure described in \cite{edge2025}.
At the beginning of the process, for identifying the most pertinent contextual information (i.e. entities and relations from the graph) corresponding to a given query, an edge-based semantic search mechanism is employed.
\item a supplementary \textit{answer evaluation} module is incorporated to assess the quality of the generated response against a set of predefined criteria. If the response satisfies the established quality thresholds, the processing pipeline terminates, and the result is delivered to the user. Conversely, if the quality requirements are not met, the process reverts to the retrieval phase, where the edge-based semantic search is relaxed to extract a broader contextual scope from the knowledge graph, thereby enhancing the likelihood of producing an improved answer.
\item a \textit{context enrichment} module, that expands the query-relevant context by inspecting the KG and retrieving additional entities and relations that were not included in the previous context.
\item the feedback loop described in the previous step is iteratively executed until a predefined termination condition is met, ensuring that the retrieval and generation processes converge toward an optimal response quality. In our implementation, we stop after four iterations, as this proved sufficient to achieve semantically relevant responses.  
\end{itemize}

The feedback loop continuously refines the retrieval process through iterative graph exploration. Unlike traditional RAG pipelines, which rely on static retrieval from a fixed knowledge base, our system adapts the context at each iteration based on the quality of the previous answers. This mechanism enables a more context-aware reasoning process, by also maximizing the resources used, as we try to explore different graph depths only when needed. 

In the following subsections we describe each key component  of our system in detail.


\subsection{Question processing}

The process starts with a question launched by the user. Optionally, the question may be processed by a Named Entity Recognition (NER) module to extract specific named entities mentioned within it. Subsequently, the question is embedded into a vector space that is compatible with the LLM and suitable for the semantic search operation performed over the KG in the following step.
Experimental results showed that NER improves overall system effectiveness, so it was retained in the KGiRAG architecture.

\subsection{Edge-based semantic search}

Each iteration begins with identifying the relevant graph context to a given question. Given that graph communities are available in the format supplied by the relevant processing step of \cite{edge2025},  by performing a cosine similarity search between question and graph community embeddings, the system retrieves textual units and community summaries that are semantically aligned with the question. Entities and relations retrieved in this step will constitute the fringe used by the context enrichment module in the subsequent iterations to navigate deeper in the KG.  

 
 
\subsection{Prompt generation}

 All retrieved textual and relational evidence is collected in the context and re-ranked according to its semantic relevance to the question. Duplicate segments are removed, and the results are categorized into two groups: (1) structured factual triples and (2) unstructured text units. 
The top-ranked segments are concatenated into a structured prompt containing both the question and its supporting context, ensuring that the LLM receives balanced factual and semantic information.
 Next, the generated prompt is passed to an LLM, which produces an answer.

The prompt (presented in figure \ref{fig:prompt}) is composed of the following three elements: 
\begin{enumerate}
\item the user’s question, 
\item the contextual information retrieved from the KG, specifically the section labeled Relevant Text Units and 
\item a brief instruction describing how the answer should be formulated, e.g., concise and limited strictly to the provided context, with an explicit statement if the answer is not present.
\end{enumerate} 

\begin{figure}[!tb]
\scriptsize
\begin{tabular}{|p{7.11cm}|}
\hline
You are a helpful question-answering assistant. \textbackslash n \\
 Question: $\{$question$\}$ \textbackslash n \\
 Context: $\{$context$\}$ \textbackslash n \\
 Instructions: \textbackslash n \\
- Use the information provided in the context to answer the question. \textbackslash n \\
- Pay special attention to the 'Relevant text units', which contain structured facts from the knowledge graph. \textbackslash n \\
- Be direct and concise in your answer. \textbackslash n \\
- If the context truly lacks the information needed to answer, say: 'The context does not provide enough information to answer this question.' \textbackslash n \\
- Do not fabricate information that is not supported by the context. \\
Answer: \textbackslash n \\
	\hline
\end{tabular}
\caption{The prompt template supplied to the LLM.}%
\label{fig:prompt}%
\end{figure}

\subsection{Answer generation and evaluation}

 The output is automatically evaluated using a suite of metrics assessing the quality of the produced text with respect to the supplied context. These metrics are divided in two categories:
\begin{itemize}
\item assessing the quality of the produced text, viewed as the output of a summarization task against the information present in the context. Here we mention standard metrics like  BERTScore \cite{bertscore} or cosine similarity
\item LLM-as-a-judge \cite{llm-as-a-judge} metrics, where the LLM's response and the context is sent back to another LLM which is then asked to compute an assessment. We used the following LLM-as-a-judge metrics: faithfulness, completeness, and relevance, defined according to \cite{ragas}, which we computed directly, given that all necessary inputs to the metric formulas were available during experimentation
\end{itemize}

These computed evaluations are logged at each iteration, enabling us to discover whether the quality of the responses improves or not and to report the obtained results. 

If the  answer quality stays below a pre-established threshold, the system re-enters the retrieval phase, by getting into the context enrichment module. This creates a closed-loop feedback mechanism that continues until convergence or a maximum number of iterations is reached.

In our case, the quality standards (from the algorithm described in fig.~\ref{fig:pseudocode}) for a response to be accepted are that both faithfulness and completeness should be at least 0.8. 
Furthermore, to ensure that the processing pipeline terminates, the response is retrieved after a maximum of four iterations, regardless of its quality. 

\subsection{Context enrichment}

If the response quality does not meet the established threshold, the process enters the context enrichment module. Here we start from the fringe in the KG composed by the entities already selected in the context and starting with them, we navigate one edge more in the graph in the hope of discovering newer entities and relations not yet selected in the previous context. We enlarge the context with the newly discovered entities and relations. 

\section{\uppercase{Experiments}}
\label{seq:experiments}

In this section we present our experimentation setup in order to evaluate the effectiveness of the system described in section~\ref{seq:methodology}. We start by presenting the dataset and then, we introduce our implementation decision, which, alongside the obtained results, could be consulted at \url{https://github.com/IacobIsabelaIE/IterativeGraphRAG}. 

Experiments were conducted on a computer with an AMD Ryzen 7 6800HS Creator Edition (3.20 GHz) CPU and 16Gb RAM.

For the LLM we used Mixtral-8x7B-Instruct-v0.1\footnote{\url{https://huggingface.co/mistralai/Mixtral-8x7B-Instruct-v0.1}}, benefiting from the fact that it is open source, LLM Arena reports good performance of Mixtral compared with similar open source models, and its size is manageable with standard computing resources.  
For the vector embedding space we selected all-MiniLM-L6-v2\footnote{\url{https://huggingface.co/sentence-transformers/all-MiniLM-L6-v2}}~\cite{sentence-bert}, which is a sentence transformer suitable for the semantic similarity computations performed in the \textit{Edge-based semantic search} component and also, it works well with general LLMs. 

\subsection{Dataset}

For this study, we adopt the \textbf{HotPotQA} dataset as provided in Microsoft’s \textit{GraphRAG Benchmarking Datasets} repository\footnote{\url{https://github.com/microsoft/graphrag-benchmarking-datasets/tree/main/data}}. The repository includes standardized data splits, annotations, and supporting facts used to evaluate graph-augmented language model architectures. HotPotQA was also used in \cite{edge2025} and allows us to compare our results with theirs and reuse their logic for constructing the KG and the graph communities. 


HotPotQA is a multi-hop question answering benchmark built over Wikipedia, originally comprising over 100,000 question–answer pairs with gold supporting paragraphs. Each question typically requires reasoning across multiple documents, making it particularly suitable for evaluating retrieval and reasoning strategies in LLMs. For this work, we utilize the filtered subset of approximately 5,000 questions released in Microsoft’s official repository, which has been preprocessed and curated specifically for GraphRAG evaluation.

To construct structured retrieval contexts, we transform the supporting paragraphs and annotated facts into a knowledge graph representation. While \cite{edge2025} used gpt-4-turbo OpenAI LLM, we re-implemented their logic for KG construction and graph community detection using Mixtral. 
The knowledge graph constructed in this way serves as input to our iterative retrieval pipeline, enabling a direct comparison between the \textit{static} retrieval used in \cite{edge2025} and our proposed iterative feedback-based retrieval strategy.

\subsection{System configurations}

The experimental component of this study is run on four architectures, listed below.
\begin{enumerate}
\item KGiRAG plus NER: our entire iterative feedback-driven retrieval system, described in section \ref{seq:methodology}, 
\item KGiRAG: the system without the NER component. We see this system version as an ablation experiment, designed to verify the robustness of our approach
\item the Microsoft's GraphRAG~\cite{edge2025}. We employ this system  as a baseline for comparative evaluation. 
\item RARR~\cite{rarr}, used to compare our GraphRAG approach  with a method from the literature that reports strong performance on sensemaking queries. 
\end{enumerate}

For convenience, Microsoft GraphRAG and RARR were executed using GPT-4o mini\footnote{\url{https://platform.openai.com/docs/models/gpt-4o-mini}}.
 
Each of these configurations were evaluated on the same subset of 500 questions extracted from the HotPotQA data set curated by Microsoft, ensuring reproducibility and direct comparability of results.  This sample covers various knowledge domains and entity types to test the graph coverage. 

%

\subsection{Evaluation metrics}


We evaluate the retrieved responses with the following metrics: 
\begin{itemize}
\item Faithfulness~\cite{ragas}: assesses how well the generated answer adheres to factual evidence in the retrieved context.
\item Completeness~\cite{ragas}: checks whether all relevant information from the context is reflected in the answer.
\item Relevance~\cite{ragas}: computes alignment between the generated answer and the original questions.
\item BERTScore~\cite{bertscore}: computes semantic similarity between generated and the context, with the embeddings computed using a masked language model (MLM) like BERT
\item Cosine similarity: embedding-based contextual similarity between the generated answer and the ground-truth evidence
\end{itemize}

\begin{table*}[!t]
\centering
\footnotesize

\setlength{\tabcolsep}{3pt} 
\begin{tabular}{lcccc}
\toprule
\textbf{Metric} & \textbf{KGiRAG with NER} & \textbf{KGiRAG without NER} & \textbf{Microsoft GraphRAG} & \textbf{RARR} \\
\midrule
Faithfulness & \textbf{0.95 $\pm$ 0.01881} & 0.78 $\pm$ 0.03613 & \textbf{0.95 $\pm$ 0.01749} & 0.90 $\pm$ 0.02615 \\
Completeness & \textbf{0.86 $\pm$ 0.029802} & 0.69 $\pm$ 0.03663 & 0.62 $\pm$ 0.0396 & 0.41 $\pm$ 0.04145 \\
Relevance & 0.64 $\pm$ 0.042269 & \textbf{0.76 $\pm$ 0.0371} & 0.23 $\pm$ 0.03549 & 0.59 $\pm$ 0.04282 \\
BERTScore & 0.79 $\pm$ 0.001603 & \textbf{0.81 $\pm$ 0.001423} & 0.77 $\pm$ 0.001551 & 0.797 $\pm$ 0.0021 \\
Cosine Similarity & 0.325 $\pm$ 0.02692 & \textbf{0.52 $\pm$ 0.01828} & 0.075 $\pm$ 0.01831 & 0.33 $\pm$ 0.02427 \\

\bottomrule
\end{tabular}
\caption{Comparison of retrieval architectures across evaluation metrics (mean and margin of error for computing the $95\%$ confidence intervals).}
\label{tab:results}
\end{table*}

\begin{figure*}[!t]
\centering

\begin{subfigure}{0.48\textwidth}
  \includegraphics[width=\textwidth]{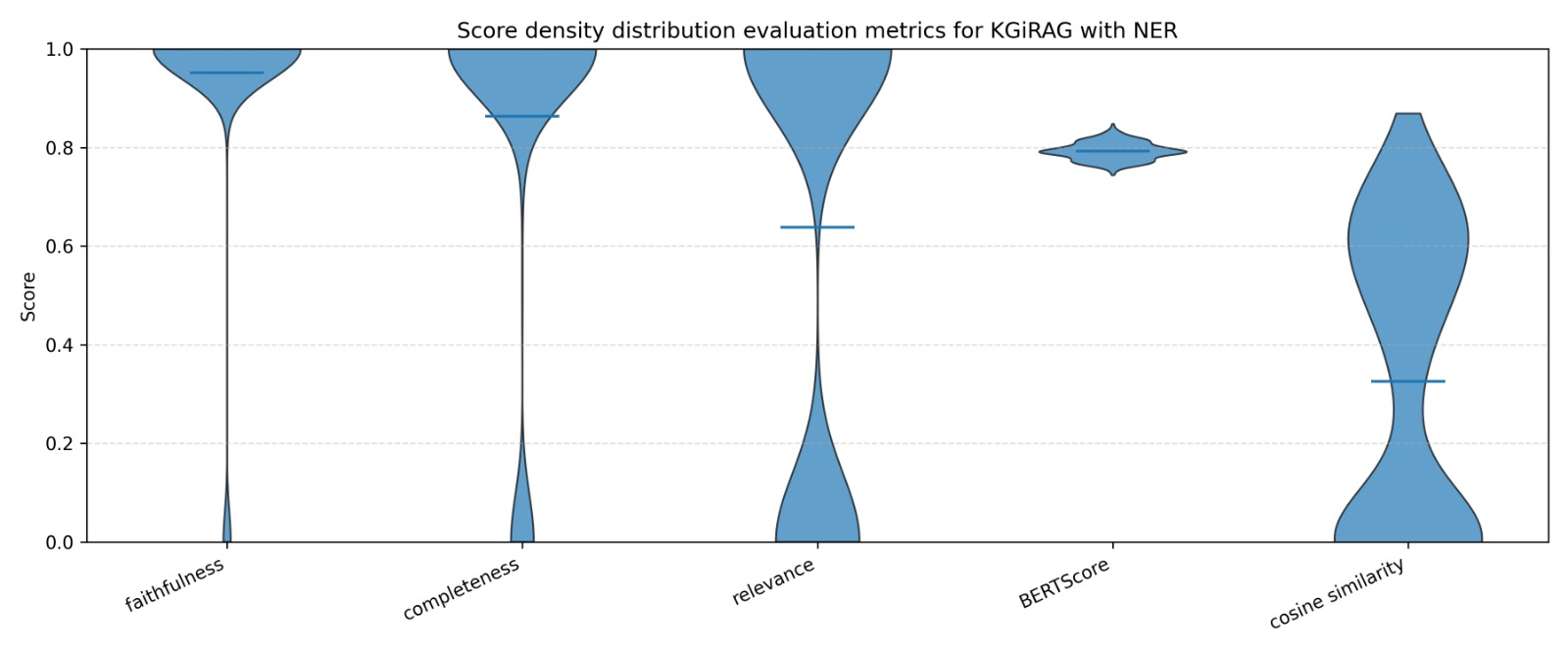}
	\caption{KGiRAG with NER}
	\label{fig:densityNER}
\end{subfigure}
\hfill
\begin{subfigure}{0.48\textwidth}
  \includegraphics[width=\textwidth]{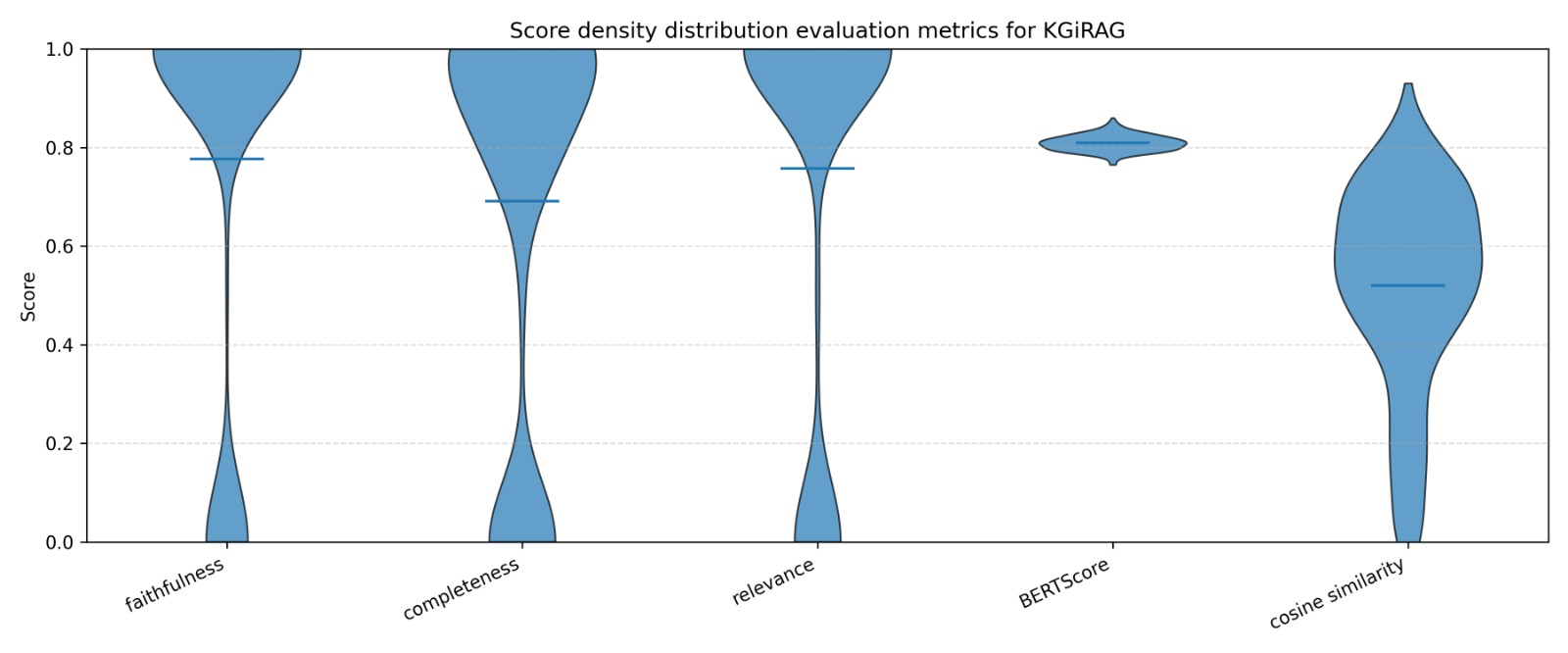}
	\caption{KGiRAG without NER}
\end{subfigure}
\vspace{0.1cm}
\begin{subfigure}{0.48\textwidth}
  \includegraphics[width=\textwidth]{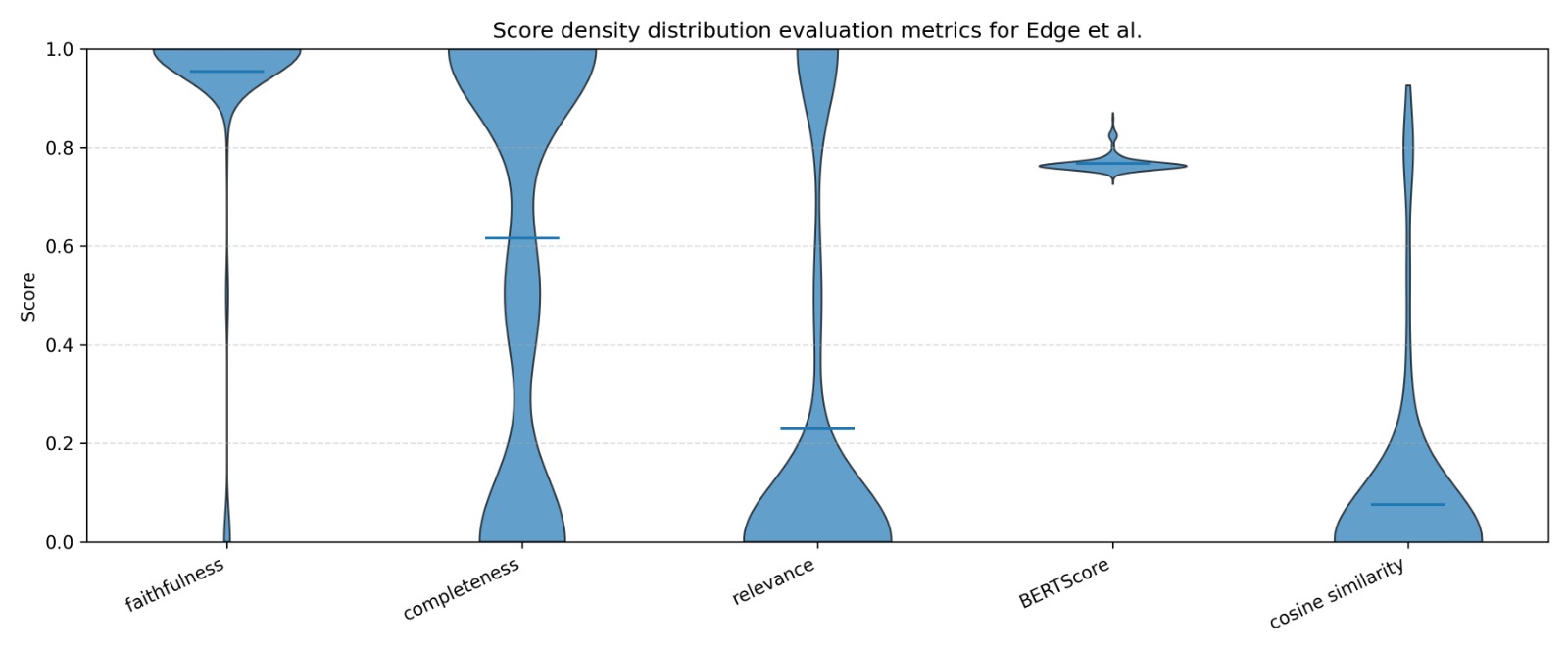}
	\caption{Microsoft GraphRAG}
	\label{fig:densityMicrosoft}
\end{subfigure}
\begin{subfigure}{0.48\textwidth}
  \includegraphics[width=\linewidth]{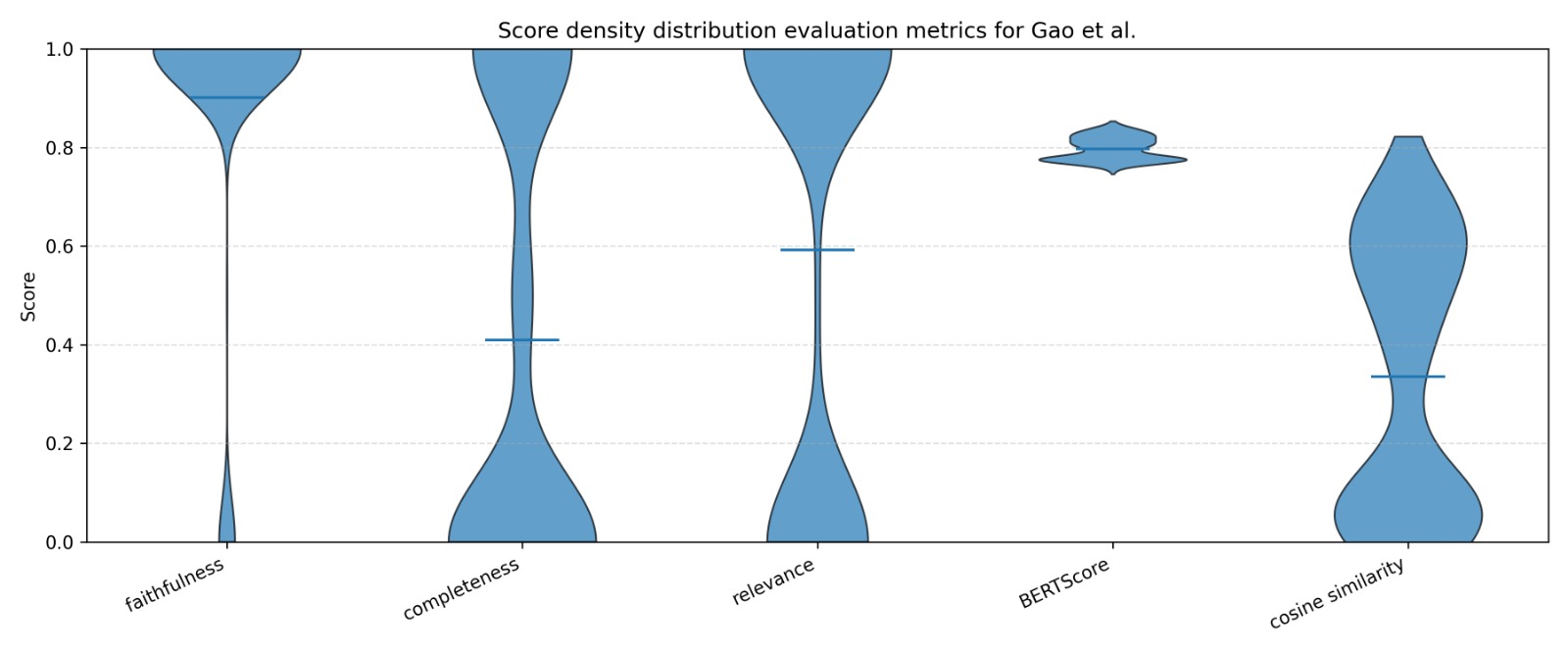}
	\caption{RARR}
\end{subfigure}
\caption{Score density plots of the evaluation metrics for the evaluated architectures}
\label{fig:scoredensity}
\end{figure*}

\section{\uppercase{Results and Discussion}}
\label{seq:results}

In this section we present our results and discuss them, in order to draw some conclusions regarding the effectiveness of our approach.

Table~\ref{tab:results} summarizes the aggregated statistics computed across all evaluated questions. Each value represents the mean of per-question scores, accompanied by the margin of error needed for computing the $95\%$ confidence intervals. Figure~\ref{fig:scoredensity} depicts the density distributions of the metrics for each evaluated system.
Together, these values provide a global picture about the effectiveness of the KGiRAG architecture. Several conclusions are drawn in what follows.

%

%

The sample size of 500 questions allows us to compute reliable values for each metric in all architectures - i.e. the margins of errors are very small, thus the expected value of each metric represents a reliable estimator for that respective metric. This allows us to conclude about a metric being superior to another using the mean values, with a $95\%$ confidence. 

We notice that the NER module being part of KGiRAG does translate into an increase in performance, confirming previous literature~\cite{ner1,ner2}. 

It seems that the leading architectures are KGiRAG including the NER module and the Microsoft GraphRAG \cite{edge2025}, both being better than RARR. We confirm previous findings in the literature \cite{graphrag,zhangsurvey} that associating a knowledge graph with the data and employing GraphRAG can improve the performance of RAG approaches when responding to sense-making queries. 


Inspecting the LLM-as-a-judge metrics, we notice that KGiRAG plus NER not only reaches the faithfulness of Microsoft GraphRAG, but also significantly improves the completeness and relevance of the responses. The semantic relevance metrics support a similar conclusion. BERTScore is slightly better, but cosine similarity is well-improved. Score density plots (figures \ref{fig:densityNER} and \ref{fig:densityMicrosoft}) display more stable distributions for KGiRAG, thus indicating a reduction in hallucination.

\begin{table}[!tb]
\centering
\footnotesize
\setlength{\tabcolsep}{3pt} 
\begin{tabular}{lcc}
\toprule
\textbf{Iteration} & \textbf{KGiRAG plus NER} & \textbf {KGiRAG} \\
\midrule
1 & 274 & 299 \\
2 & 83 & 41 \\
3 & 59 & 15 \\
4 & 84 & 145 \\
\bottomrule
\end{tabular}
\caption{Number of responses solved in each of the four iterations of the KGiRAG}
\label{tbl:iterations}
\end{table}

\begin{figure*}[!t]
  \centering
\begin{subfigure}{0.48\textwidth}
  \includegraphics[width=0.9\textwidth]{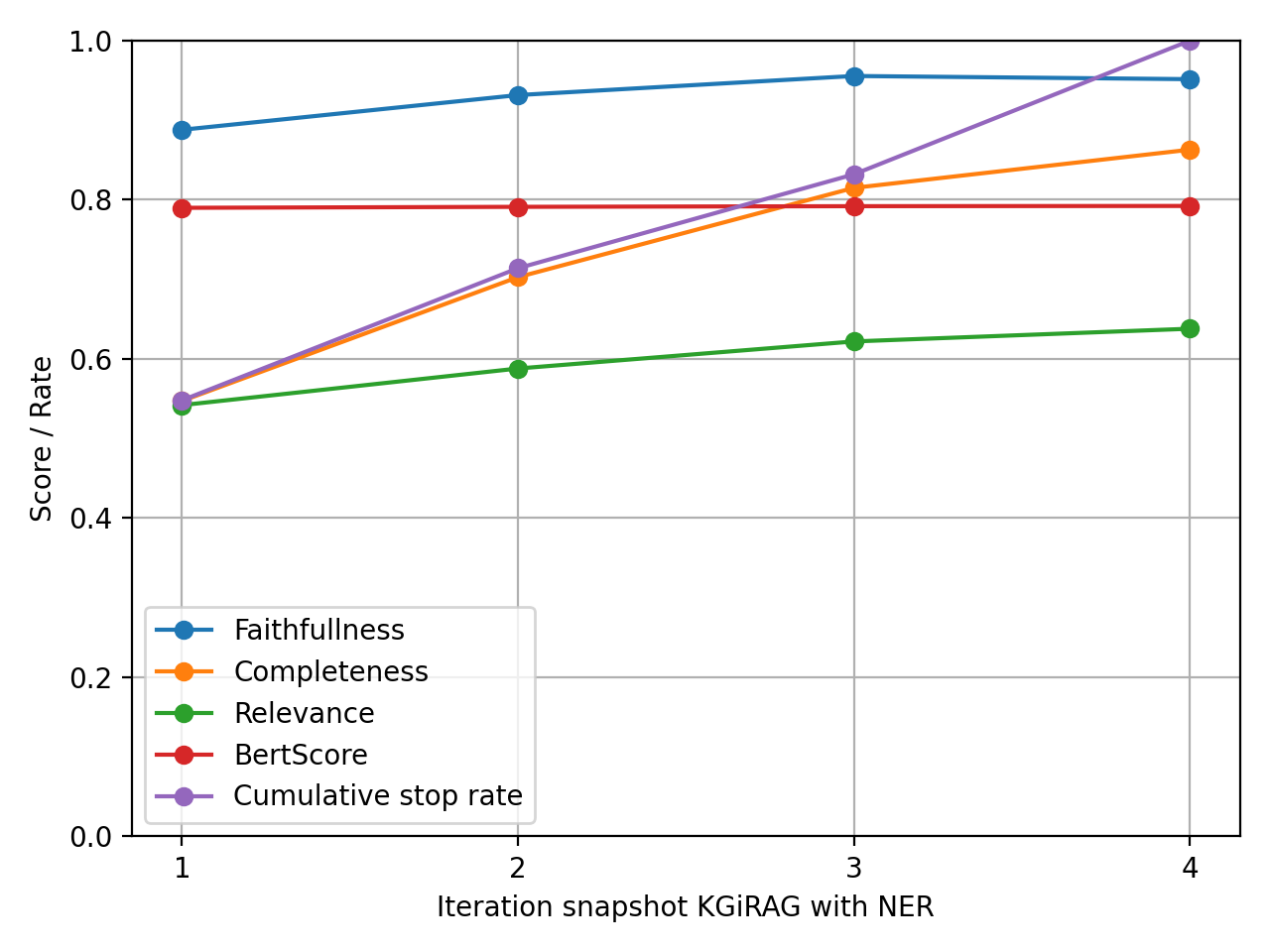}
	\caption{KGiRAG with NER}
	\label{fig:iterationNER}
\end{subfigure}
\hfill
\begin{subfigure}{0.48\textwidth}
  \includegraphics[width=0.9\textwidth]{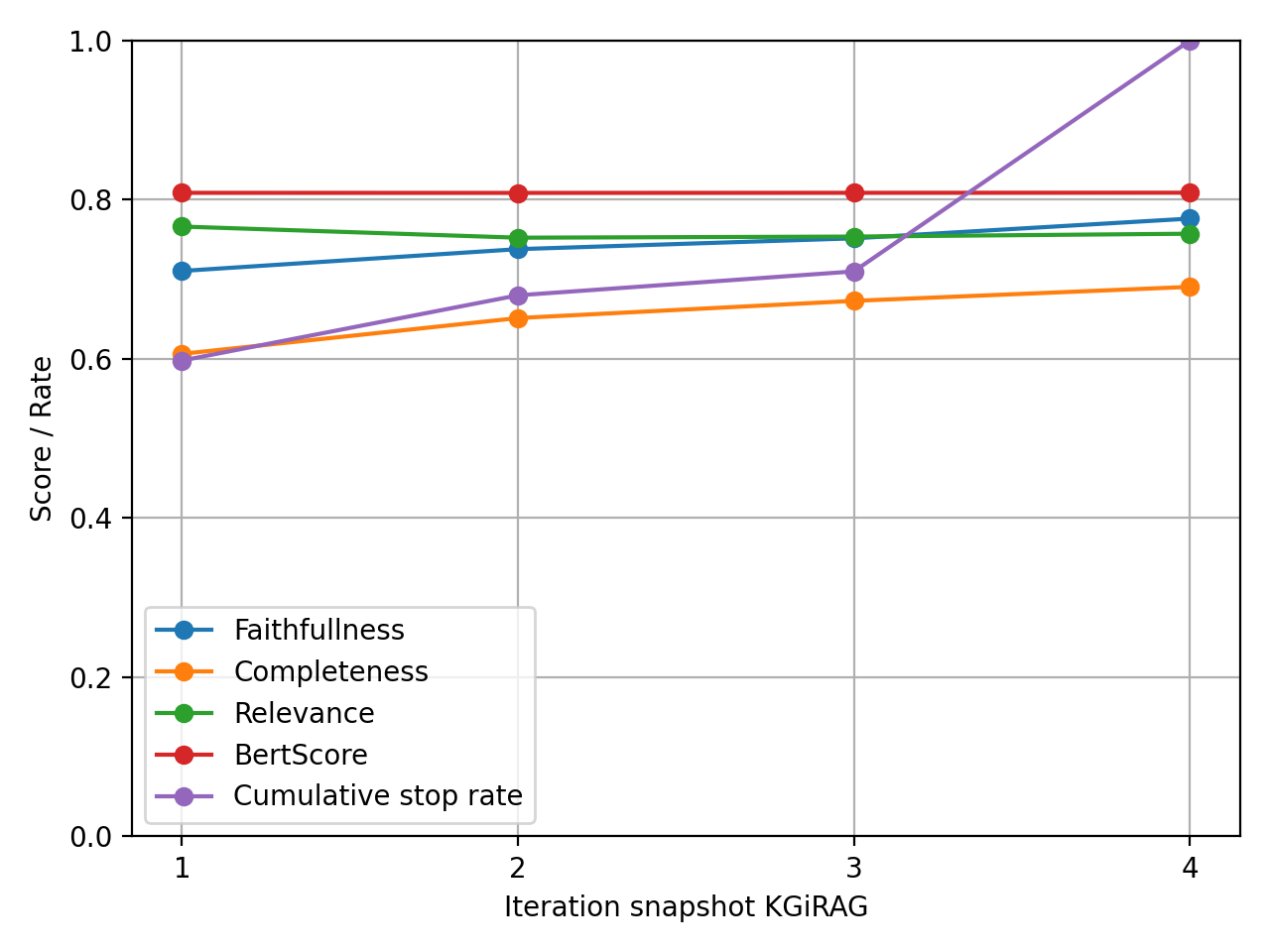}
	\caption{KGiRAG without NER}
	\label{fig:iteration}
\end{subfigure}
\caption{Metrics evolution during iterations for KGiRAG}
\label{fig:rounds}
\end{figure*}

Table \ref{tbl:iterations}  and figures \ref{fig:iterationNER} and \ref{fig:iteration} show how the iterative approach considered in KGiRAG improves the performance of the metrics through gradually expanding the query context. The quality thresholds imposed by the architecture (i.e., faithfulness and completeness greater than 0.8) are initially satisfied by responses for only about half of the queries. As the iterations proceed, an increasing number of queries achieve higher-quality responses. We further observe that the inclusion of the NER module substantially improves performance, reducing the number of unresolved queries in the final iteration by approximately~$40\%$.

From the evolution of the faithfulness metric, shown by the blue line in Figure \ref{fig:iterationNER}, we observe that it reaches a peak value of 0.95 at the third iteration. Consequently, we terminate the iterative procedure after four rounds. This stopping criterion could be replaced with alternative conditions, such as halting when faithfulness or another evaluation metric ceases to improve over multiple iterations. Whether to adopt a hard stopping rule, as in our case, or a softer, adaptive one depends on the available computational resources and on the expressiveness and completeness of the knowledge graph associated with the data.

We believe that iteratively expanding the context represents a more effective design choice than single-shot alternatives, as it better balances the trade-off between achieving high-quality responses and limiting the use of expensive computational resources. Microsoft GraphRAG~\cite{edge2025} extracts a large amount of contextual information during its initial phase, while GRAG \cite{grag} expands the knowledge graph in a query-driven and depth-bounded manner. For knowledge graphs with a high branching factor, such strategies could incur substantial computational costs and result in excessively large contexts being passed to the LLM. This may negatively affect system performance, particularly with respect to LLM-as-a-judge metrics, which are sensitive to the size of the context provided to the language model.

\section{\uppercase{Conclusion}}
\label{seq:conclusion}

In this paper, we introduce a novel iterative, feedback-driven GraphRAG architecture designed to address sensemaking queries using a large language model (LLM). Our approach aims to mitigate the LLM’s tendency to produce hallucinations, particularly in cases where sensemaking queries require comprehensive, evidence-based responses grounded in extensive domain-specific source documents.

Assuming that the underlying knowledge is formalized and stored in a knowledge graph, our findings align with prior work~\cite{graphgpt,graphrag,edge2025,grag}, confirming that integrating graph-structured data into the RAG pipeline enhances the quality of LLM-generated responses. Moreover, we demonstrate the advantages of an iterative RAG approach over a single-shot strategy, as it yields responses with higher semantic quality and improved relevance and completeness.

The results presented in this paper are preliminary. Further experimentation is necessary using additional datasets—preferably those with a well-defined ground truth—to enable a more precise evaluation of the system's performance. Moreover, comprehensive ablation studies are required to quantify the contribution of each component in the RAG pipeline to the overall answer quality. Finally, future work will involve testing with more advanced LLMs and fine-tuning pipeline parameters, including alternative criteria that determine whether a response is accepted as final.  

\section*{\uppercase{Acknowledgements}}

Isabela Iacob thanks Babe\c{s}-Bolyai University for granting her the special scholarship for scientific activity during the academic year 2024-2025.  

\bibliographystyle{apalike}
{\small
\bibliography{sample}}

@article{zhangsurvey,
  author       = {Qinggang Zhang and
                  Shengyuan Chen and
                  Yuanchen Bei and
                  Zheng Yuan and
                  Huachi Zhou and
                  Zijin Hong and
                  Junnan Dong and
                  Hao Chen and
                  Yi Chang and
                  Xiao Huang},
  title        = {{A Survey of Graph Retrieval-Augmented Generation for Customized Large
                  Language Models}},
  journal      = {CoRR},
  volume       = {abs/2501.13958},
  year         = {2025},
  doi          = {10.48550/ARXIV.2501.13958},
}

@inproceedings{sentence-bert,
  author       = {Nils Reimers and
                  Iryna Gurevych},
  editor       = {Kentaro Inui and
                  Jing Jiang and
                  Vincent Ng and
                  Xiaojun Wan},
  title        = {{Sentence-BERT: Sentence Embeddings using Siamese BERT-Networks}},
  booktitle    = {Proc. of the 2019 Conf. on Empirical Methods in Natural
                  Language Processing and the 9th Intl. Joint Conf. on
                  Natural Language Processing, {EMNLP-IJCNLP} 2019, Hong Kong, China },
  pages        = {3980--3990},
  publisher    = {ACL},
  year         = {2019},
  doi          = {10.18653/V1/D19-1410},
}

@inproceedings{ner1,
  author       = {Jinyoung Kim and
                  Dayoon Ko and
                  Gunhee Kim},
  editor       = {Yaser Al{-}Onaizan and
                  Mohit Bansal and
                  Yun{-}Nung Chen},
  title        = {{DynamicER: Resolving Emerging Mentions to Dynamic Entities for RAG}},
  booktitle    = {Proc. of the 2024 Conf. on Empirical Methods in Natural
                  Language Processing, {EMNLP} 2024, Miami, USA},
  pages        = {13752--13770},
  publisher    = {ACL},
  year         = {2024},
  doi          = {10.18653/V1/2024.EMNLP-MAIN.762},
}

@inproceedings{ner2,
    title = {{Entity Retrieval for Answering Entity-Centric Questions}},
    author = "Shavarani, Hassan  and
      Sarkar, Anoop",
    editor = "Shi, Weijia  and
      Yu, Wenhao  and
      Asai, Akari  and
      Jiang, Meng  and
      Durrett, Greg  and
      Hajishirzi, Hannaneh  and
      Zettlemoyer, Luke",
    booktitle = "Proc. of the 4th Intl. Workshop on Knowledge-Augmented Methods for Natural Language Processing",
    year = "2025",
    publisher = "ACL",
    doi = "10.18653/v1/2025.knowledgenlp-1.1",
    pages = "1--17",
  }

@inproceedings{kirag,
  author       = {Jinyuan Fang and
                  Zaiqiao Meng and
                  Craig MacDonald},
  editor       = {Wanxiang Che and
                  Joyce Nabende and
                  Ekaterina Shutova and
                  Mohammad Taher Pilehvar},
  title        = {{KiRAG: Knowledge-Driven Iterative Retriever for Enhancing Retrieval-Augmented
                  Generation}},
  booktitle    = {Proc. of the 63rd Annual Meeting of the ACL
                   (Volume 1: Long Papers), {ACL} 2025, Vienna, Austria},
  pages        = {18969--18985},
  publisher    = {ACL},
  year         = {2025},
}

@inproceedings{rarr,
  author       = {Luyu Gao and
                  Zhuyun Dai and
                  Panupong Pasupat and
                  Anthony Chen and
                  Arun Tejasvi Chaganty and
                  Yicheng Fan and
                  Vincent Y. Zhao and
                  Ni Lao and
                  Hongrae Lee and
                  Da{-}Cheng Juan and
                  Kelvin Guu},
  editor       = {Anna Rogers and
                  Jordan L. Boyd{-}Graber and
                  Naoaki Okazaki},
  title        = {{RARR: Researching and Revising What Language Models Say, Using Language
                  Models}},
  booktitle    = {Proc. of the 61st Annual Meeting of the ACL
                  (Volume 1: Long Papers), {ACL} 2023, Toronto, Canada},
  pages        = {16477--16508},
  publisher    = {ACL},
  year         = {2023},
  doi          = {10.18653/V1/2023.ACL-LONG.910},
}

@article{graphrag,
  author       = {Haoyu Han and
                  Yu Wang and
                  Harry Shomer and
                  Kai Guo and
                  Jiayuan Ding and
                  Yongjia Lei and
                  Mahantesh Halappanavar and
                  Ryan A. Rossi and
                  Subhabrata Mukherjee and
                  Xianfeng Tang and
                  Qi He and
                  Zhigang Hua and
                  Bo Long and
                  Tong Zhao and
                  Neil Shah and
                  Amin Javari and
                  Yinglong Xia and
                  Jiliang Tang},
  title        = {{Retrieval-Augmented Generation with Graphs (GraphRAG)}},
  journal      = {CoRR},
  volume       = {abs/2501.00309},
  year         = {2025},
  doi          = {10.48550/ARXIV.2501.00309},
  eprinttype    = {arXiv},
}

@inproceedings{grag,
  author       = {Yuntong Hu and
                  Zhihan Lei and
                  Zheng Zhang and
                  Bo Pan and
                  Chen Ling and
                  Liang Zhao},
  editor       = {Luis Chiruzzo and
                  Alan Ritter and
                  Lu Wang},
  title        = {{GRAG: Graph Retrieval-Augmented Generation}},
  booktitle    = {Findings of the ACL: {NAACL}
                  2025, Albuquerque, New Mexico, USA},
  pages        = {4145--4157},
  publisher    = {ACL},
  year         = {2025},
  doi          = {10.18653/V1/2025.FINDINGS-NAACL.232},
}

@article{hallucination,
  author       = {Lei Huang and
                  Weijiang Yu and
                  Weitao Ma and
                  Weihong Zhong and
                  Zhangyin Feng and
                  Haotian Wang and
                  Qianglong Chen and
                  Weihua Peng and
                  Xiaocheng Feng and
                  Bing Qin and
                  Ting Liu},
  title        = {{A Survey on Hallucination in Large Language Models: Principles, Taxonomy,
                  Challenges, and Open Questions}},
  journal      = {{ACM} Trans. Inf. Syst.},
  volume       = {43},
  number       = {2},
  pages        = {42:1--42:55},
  year         = {2025},
  doi          = {10.1145/3703155},
}

@inproceedings{kandpal,
  author       = {Nikhil Kandpal and
                  Haikang Deng and
                  Adam Roberts and
                  Eric Wallace and
                  Colin Raffel},
  editor       = {Andreas Krause and
                  Emma Brunskill and
                  Kyunghyun Cho and
                  Barbara Engelhardt and
                  Sivan Sabato and
                  Jonathan Scarlett},
  title        = {{Large Language Models Struggle to Learn Long-Tail Knowledge}},
  booktitle    = {Intl. Conf. on Machine Learning, {ICML} 2023, Hawaii, {USA}},
  series       = {Proceedings of Machine Learning Research},
  volume       = {202},
  pages        = {15696--15707},
  publisher    = {{PMLR}},
  year         = {2023},
  url          = {https://proceedings.mlr.press/v202/kandpal23a.html},
}

@inproceedings{bertscore,
  author       = {Tianyi Zhang and
                  Varsha Kishore and
                  Felix Wu and
                  Kilian Q. Weinberger and
                  Yoav Artzi},
  title        = {{BERTScore: Evaluating Text Generation with BERT}},
  booktitle    = {8th Intl. Conf. on Learning Representations, {ICLR} 2020,
                  Addis Ababa, Ethiopia},
  publisher    = {OpenReview.net},
  year         = {2020},
  url          = {https://openreview.net/forum?id=SkeHuCVFDr},
}

@inproceedings{ragas,
  author       = {Shahul ES and
                  Jithin James and
                  Luis Espinosa Anke and
                  Steven Schockaert},
  editor       = {Nikolaos Aletras and
                  Orph{\'{e}}e De Clercq},
  title        = {{RAGAs: Automated Evaluation of Retrieval Augmented Generation}},
  booktitle    = {Proc. of the 18th Conf. of the European Chapter of the
                  ACL, {EACL} 2024 - System Demonstrations,
                  St. Julians, Malta},
  pages        = {150--158},
  publisher    = {ACL},
  year         = {2024},
  url          = {https://aclanthology.org/2024.eacl-demo.16},
}

@inproceedings{llm-as-a-judge,
  author       = {Lianmin Zheng and
                  Wei{-}Lin Chiang and
                  Ying Sheng and
                  Siyuan Zhuang and
                  Zhanghao Wu and
                  Yonghao Zhuang and
                  Zi Lin and
                  Zhuohan Li and
                  Dacheng Li and
                  Eric P. Xing and
                  Hao Zhang and
                  Joseph E. Gonzalez and
                  Ion Stoica},
  editor       = {Alice Oh and
                  Tristan Naumann and
                  Amir Globerson and
                  Kate Saenko and
                  Moritz Hardt and
                  Sergey Levine},
  title        = {Judging {LLM-as-a-Judge} with {MT-Bench} and {Chatbot Arena}},
  booktitle    = {Advances in Neural Information Processing Systems 36, NeurIPS 2023, New Orleans,
                  LA, USA},
  year         = {2023},
  url          = {http://papers.nips.cc/paper\_files/paper/2023/hash/91f18a1287b398d378ef22505bf41832-Abstract-Datasets\_and\_Benchmarks.html},
}

@inproceedings{hotpotqa,
  author       = {Zhilin Yang and
                  Peng Qi and
                  Saizheng Zhang and
                  Yoshua Bengio and
                  William W. Cohen and
                  Ruslan Salakhutdinov and
                  Christopher D. Manning},
  editor       = {Ellen Riloff and
                  David Chiang and
                  Julia Hockenmaier and
                  Jun'ichi Tsujii},
  title        = {{HotpotQA: {A} Dataset for Diverse, Explainable Multi-hop Question
                  Answering}},
  booktitle    = {Proc. of the 2018 Conf. on Empirical Methods in Natural
                  Language Processing, Brussels, Belgium},
  pages        = {2369--2380},
  publisher    = {ACL},
  year         = {2018},
  doi          = {10.18653/V1/D18-1259},
}

@article{edge2025,
  author       = {Darren Edge and
                  Ha Trinh and
                  Newman Cheng and
                  Joshua Bradley and
                  Alex Chao and
                  Apurva Mody and
                  Steven Truitt and
                  Jonathan Larson},
  title        = {{From Local to Global: {A} Graph {RAG} Approach to Query-Focused Summarization}},
  journal      = {CoRR},
  volume       = {abs/2404.16130},
  year         = {2024},
  doi          = {10.48550/ARXIV.2404.16130},
}

@inproceedings{graphgpt,
  author       = {Jiabin Tang and
                  Yuhao Yang and
                  Wei Wei and
                  Lei Shi and
                  Lixin Su and
                  Suqi Cheng and
                  Dawei Yin and
                  Chao Huang},
  editor       = {Grace Hui Yang and
                  Hongning Wang and
                  Sam Han and
                  Claudia Hauff and
                  Guido Zuccon and
                  Yi Zhang},
  title        = {{GraphGPT: Graph Instruction Tuning for Large Language Models}},
  booktitle    = {Proc. of the 47th Intl. {ACM} {SIGIR} Conf. on
                  Research and Development in Information Retrieval, {SIGIR} 2024, Washington
                  DC, USA},
  pages        = {491--500},
  publisher    = {{ACM}},
  year         = {2024},
  doi          = {10.1145/3626772.3657775},
}

@inproceedings{lewis2021,
  author       = {Patrick Lewis and
                  Ethan Perez and
                  Aleksandra Piktus and
                  Fabio Petroni and
                  Vladimir Karpukhin and
                  Naman Goyal and
                  Heinrich K{\"{u}}ttler and
                  Mike Lewis and
                  Wen{-}tau Yih and
                  Tim Rockt{\"{a}}schel and
                  Sebastian Riedel and
                  Douwe Kiela},
  editor       = {Hugo Larochelle and
                  Marc'Aurelio Ranzato and
                  Raia Hadsell and
                  Maria{-}Florina Balcan and
                  Hsuan{-}Tien Lin},
  title        = {{Retrieval-Augmented Generation for Knowledge-Intensive NLP Tasks}},
  booktitle    = {Advances in Neural Information Processing Systems 33, NeurIPS 2020, virtual},
  year         = {2020},
  url          = {https://proceedings.neurips.cc/paper/2020/hash/6b493230205f780e1bc26945df7481e5-Abstract.html},
}

\end{document}